\documentclass[11pt,fleqn]{article}   

\usepackage{latexsym,bm}
\usepackage{amssymb,amsfonts}
\usepackage{amsmath}   

\catcode`\@=11
\@addtoreset{equation}{section}
\def\theequation{\arabic{section}.\arabic{equation}}

\def\appendix{\renewcommand{\thesection}{\Alph{section}}\setcounter{section}{0}
              \renewcommand{\theequation}
            {\mbox{\Alph{section}.\arabic{equation}}}\setcounter{equation}{0}}

\newcommand{\ca}[1]{{\cal #1}}         
\def\beq{\begin{eqnarray}}
\def\eeq{\end{eqnarray}}
\def\at{\left(}\def\aq{\left[}
\def\ct{\right)}
\def\cq{\right]}

\def\R{{\hbox{{\rm I}\kern-.2em\hbox{\rm R}}}}   
\def\ii{\infty}                                  

\def\de{\delta}

\def\th{\theta}

\begin{document}

\title{Some results on dynamical black holes}
\author{L.~Vanzo\thanks{vanzo@science.unitn.it}\vspace{.5cm}\\ 
Dip.~di Fisica, Universit\`a di Trento\\ 
and Ist.~Nazionale di Fisica Nucleare,\\
Gruppo Collegato di Trento,\\ Italia}

\date{}
\maketitle

\begin{abstract}
We give indications that outer future trapping horizons 
play a role in the particular semi-classical instability of an
evolving black hole that produces the Hawking's radiation. These are
obtained  with the use of the Hamilton-Jacobi tunneling method. It
automatically selects one special expression for the surface gravity
of a changing horizon, the one defined a decade ago by Hayward using
Kodama's theory of spherically symmetric gravitational fields. The
method also applies to point masses embedded in an expanding universe
and to general, spherically symmetric black holes. The local
surface gravity solves a puzzle concerning the charged stringy
black holes, namely that it vanishes in the
extremal limit, whereas the Killing global gravity does not.  
\end{abstract}

\thanks{Devoted to Prof.~I.~Brevik on the occasion of his 70th
  birthday.}

\section{Introduction}

It has long been felt that the usual semi-classical treatment of
stationary black holes (abbrev. BHs) should be extended to cover at
least slowly changing, or evolving black holes. By this expression we
mean black holes that can still be described in terms of few
multiplole moments such as mass, angular momentum and the charges
associated to local gauge symmetries, except that the parameters  
and the causal structure change with time either because
matter and gravitational radiation fall in, or because there operate 
a Hawking's process of quantum evaporation or finally because the hole
is actually immersed in a slowly expanding universe. A technical
definition of a ``slowly varying BH'' can be given in some cases, an
example being the Booth-Fairhurst slowly evolving horizon,
but in general it depends on the actual physical processes involved.
For example, in the case of Hawking's evaporation, conditions for
slowness in the presence of a near-horizon viscous fluid have been
given by Brevik \cite{Brevik:1999bj} in an interesting attempt to
generalize 't Hooft's model of the self-screening Hawking atmosphere
(quantum corrections to this model can be found in
\cite{Nojiri:1999pm}). In general it is understood that the 
black hole temperature is to be be much smaller than the Planck mass, or
equivalently the mass $M\gg M_{P}=G^{-1/2}\sim10^{19}$ Gev, while in
order to study the effects of the expansion the Hubble rate $H^{-1}$
should dominate over the black hole emission/absorption rate.

One surprising aspect of the semi-classical results obtained so far,
is that the radiation caused by  
the changing metric of the collapsing star approaches a steady
outgoing flux for large times, implying a drastic violation of energy
conservation if one neglects the back reaction of the quantum radiation
on the structure of spacetime. But the back reaction problem
has not been solved yet in a satisfactory way. As pointed out by
Fredenhagen and Haag long ago \cite{Fredenhagen:1989kr}, if the back
reaction is taken into account by letting the mass of the black hole
to change with time, then the radiation will possibly originate from
the \emph{surface} of the black hole at all times after its formation. 

This poses the question: \emph{what is and where is the surface of a
dynamical black hole}? This issue always baffled scientists from the very
beginning and produced several reactions during the nineties, which
eventually culminated with the notion of outer trapping horizons by
Hayward\cite{hayw94} and the isolated \cite{Ashtekar:1998sp} and
dynamical horizons of Ashtekar and Krishnan
\cite{Ashtekar:2003hk,ash-kris02} (a fine review is in
\cite{Ashtekar:2004cn}). Thus one 
is concerned to show, in the first place, what kind of a surface a
dynamical horizon can be and also which definition can capture a
useful local notion of such a surface, and then what sort
of instability, if any, really occurs near the horizon of the changing
black hole. This question was non trivial since a changing horizon is
typically embedded in a dynamical space-time and it is not
even expected to be a null hypersurface, although it is still one of
infinite red shift.

We analyze this question for a class of dynamical black hole
solutions that was inspired by problems not 
directly related to black hole physics, although these were
subsequently reconsidered in the light of the black hole back reaction 
problem in the early Eighties. The metrics we shall consider are the
Vaidya radiating metric \cite{vaidya}, as revisited by
J.~Bardeen \cite{bardeen} and J.~York \cite{york}, together with what
really is a fake dynamical black holes, the McVittie solution
representing, in author's mind, a point mass in
cosmology \cite{mcv}. We shall indicate how the results can be
extended to all dynamical, spherically symmetric solutions admitting
a possibly dynamical future outer trapping horizon.

\section{Horizons}

After the time lasting textbook definition of the event horizon
(abbr.~EH) to be found in the Hawking \& Ellis renowned book
\cite{HE73}, several quasi-local notions of dynamical horizons 
appeared in the 
literature (a nice review is in \cite{Gourgoulhon:2008pu})
, perhaps starting with the perfect horizons\footnote{These
are null hypersurfaces whose rays have zero expansion and intersect
space-like hypersurfaces in compact sets. All stationary horizons
  are perfect, but the converse is not true.} of H\'aji$\check{c}$ek
\cite{hajicek} and the apparent horizons (AHs, boundaries of
trapped \(3\)-dimensional space-like regions) of Hawking-Ellis
themselves. But the former only applied to equilibrium BHs and the
existence of the latter is tie to a partial Cauchy surface so it
represents only a ``localization in time''. Moreover it has proven not
possible to formulate thermodynamical laws for AHs akin to those
holding for event horizons.  

The first succesfull attempt to go beyond the limitations imposed
either by the istantaneous character of the apparent horizons or by
the global, teleological nature of the event horizons is due to
S.\ Hayward. His concept of a future outer trapping horizon
(FOTH) then evolved either into some less constrained definition, like
the Ashtekar-Krishnan dynamical horizons (DH), or some specialization
like the Booth-Fairhurst slowly evolving FOTH \cite{bf04}; so an
updated (but perhaps partial) list of locally or quasi-locally defined
horizons would contain: 
\begin{itemize}
\item[(a)] Trapping horizons (Hayward \cite{hayw94})
\item[(b)] Dynamical horizons (Ashtekar \& Krishnan
 \cite{ash-kris02,Ashtekar:2003hk})
\item[(c)] Non expanding and perfect horizons (H\'aji$\check{c}$ek
 \cite{hajicek}) 
\item[(d)] Isolated and weakly isolated horizons (Ashtekar et al.\
  \cite{Ashtekar:1998sp}) 
\item[(e)] Slowly evolving horizons (Booth \& Fairhurst \cite{bf04})
\end{itemize}
In contrast to the old fashioned apparent horizons, these newly defined
horizons 
do not require a space-like hypersurface, no notion of interior and
exterior and no conditions referring to infinity (all are non local
conditions). Moreover they are not teleological and, given a solution
of Einstein equations, one can find whether they exist
by purely local computations. Finally, unlike EHs they are related to
regions endowed by strong gravitational fields and absent in weak
field regions.  

All quasi-local horizons rely on the local concept of trapped
(marginally trapped) surface:
this is a space-like closed $2$-manifold $\bm{S}$ such that
$\theta_{(\ell)}\theta_{(n)}>0$, where $\ell$, $n$ are the
future-directed null normals to $\bm{S}$, normalized to $\ell\cdot
 n=-1$, and $\theta_{(\ell)}$, $\theta_{(n)}$ are the respective
expansion scalars. We write the induced metric on each $\bm{S}$ in the 
form 
\beq\label{qab}
q_{ab}=g_{ab}+\ell_an_b+\ell_bn_a
\eeq
and put $q^{ab}=g^{ab}+\ell^an^b+\ell^bn^a$, not an inverse. Then $q^a_b$
is the projection tensor to $T_*(\bm{S})$, the tangent 
space to $\bm{S}$. To cover BHs rather than white holes it is further
assumed that both expansions are negative (non positive). 

The most important quantities associated
with the null vector fields $\ell$ and $n$ are the projected tensor
fields $\Theta_{ab}=q_n^aq_m^b\nabla_al_b$ and
$\Phi_{ab}=q_n^aq_m^b\nabla_an_b$ and their 
decomposition into symmetric, anti-symmetric and trace part. Their
twists are zero  since they are normal to $\bm{S}$. Finally, the
expansions  are given by 
\beq\label{exp}
\th_{(\ell)}=q^{ab}\nabla_a\ell_b, \qquad \th_{(n)}=q^{ab}\nabla_an_b 
\eeq
Let us
describe the listed horizons in turn, adding comments where it
seems appropriate. A black triangle down $\blacktriangledown$ will
close the definitions.

\noindent \textbf{Future Outer Trapping Horizon:} A future outer
trapping horizon (FOTH) is a smooth three-dimensional sub-manifold $H$
of space-time which is foliated by closed space-like two-manifolds
$\bm{S}_t$, $t\in\R$, with future-directed null normals $\ell$ and $n$
such that (i) the expansion $\theta_{(\ell)}$ of the null normal $\ell$
vanishes, (ii) the expansion $\theta_{(n)}$ of $n$ is negative and (iii)  
$\ca L_{n}\theta_{(\ell)} < 0$. $\blacktriangledown$

Condition (i) requires strong fields since certainly $\th_{(\ell)}>0$ in
weak fields. Condition (ii) is related to the idea that $H$ is of the
future type (i.~e. a BH rather than a WH); (iii) says that $H$ is of
the outer type, since  a motion of $\bm{S}_t$ along $n^a$ makes it trapped. It
also distinguishes BH horizons from cosmological ones.

One can always found a scalar field $C$ on $H$ so that
\beq\label{v}
   V^{a} = \ell^{a} - C n^{a} \quad \mbox{and} \quad
   N_{a} = \ell_{a} + C n_{a} \, , 
\eeq
are respectively tangent and normal to the horizon. Note that
$V \cdot V = - N\cdot N = 2C$. Hayward
\cite{hayw94} has shown that if the null energy condition holds, then
$C \ge 0$ on a FOTH.  Thus, the horizon must be either space-like or
null, being null iff the shear $\sigma^{(\ell)}_{ab}$ as well as
$T_{ab}\ell^a\ell^b$ both vanish across $H$. Intuitively, $H$ is
space-like in the dynamical regime where gravitational radiation and
matter are pouring into it and is null when it reaches equilibrium.

The second law of trapping horizon mechanics follows quite easily. If
$\sqrt{q}$ is the area element corresponding to the metric $q_{ab}$ on
the cross-sections, then  
\begin{equation}\label{metdetder}
   \ca L_{V} \sqrt{q} = - C \theta_{(n)}\sqrt{q} \, .  
\end{equation}
By definition $\theta_{(n)}$ is negative and we have just seen that
$C$ is non-negative, so we obtain the local second law: If the
null energy condition holds, then the area element $\sqrt{q}$
of a FOTH is non-decreasing along the horizon. 
Integrating over $\bm{S}_t$, the same law applies to the total area of
the horizon sections. It is non-decreasing and remains constant if and
only if the horizon is null. 

\noindent \textbf{Dynamical Horizon:} a smooth three-dimensional, 
\emph{space-like} sub-manifold $H$ of space-time is a dynamical
horizon (DH) if it can be foliated by closed space-like two-manifolds
$\bm{S}_t$, with future-directed null normals $\ell$ and $n$ such that
(i) on each leaf the expansion $\theta_{(\ell)}$ of one null normal
$\ell^a$ vanishes, (ii) the expansion $\theta_{(n)}$ of the other
null normal $n$ is negative.$\blacktriangledown$

Like FOTHs, a DH is a space-time notion defined quasi-locally, it is
not relative to a space-like hypersurface, it does not refer to $\ii$,
it is not teleological. A space-like FOTH is a DH on which $\ca
L_n\theta_{(\ell)}< 0$; a DH which is also a FOTH will be called a
space-like future outer horizon (SFOTH). The DH cannot describe
equilibrium black holes since it is space-like by definition, but is
better suited to describe how a BH grow in general
relativity. Suitable analogues of the laws of black hole mechanics
hold for both FOTHs and DHs. Our main interest in the following will
be precisely for these local horizons, but for the time being we
continue our description.  

\noindent \textbf{Perfect and Non-Expanding Horizons:} a perfect
horizon is a smooth three-dimensional  
\emph{null} sub-manifold $H$ of space-time with null normal $\ell^a$
such that $\theta_{(\ell)}=0$ on $H$ and which intersect space-like
hypersurfaces in compact sets. $\blacktriangledown$ \\
If in the last clause $H$ is topologically $\R\times S^2$ and moreover
the stress tensor $T_{ab}$ is such that $-T^a_b\ell^b$ is future
causal for any future directed null normal $\ell^a$, then $H$ is
called a non-expanding horizon.$\blacktriangledown$ 

If $X$, $Y$ are tangent to a non-expanding horizon we can decompose
the covariant derivative 
\[
\nabla_XY=D_XY+N(X,Y)\ell+L(X,Y)n
\]
where $D_X$ is the projection of the vector $\nabla_XY$ onto the
spheres $\bm{S}_t$ in $H$. If $X$ is tangent to the spheres then $D_X$
is the covariant derivative of the induced metric $q_{ab}$, and if $X$
is tangent to $H$ one may regard the operator
$\widehat{\nabla}_X=D_X+N(X,\cdot)\ell$, acting on vector fields, as 
a connection on $H$. If this connection is ``time independent'' then
the geometry of $H$ is time independent too and we have Ashtekar et al.
notion of an horizon in isolation.      

\noindent \textbf{Isolated Horizon:} a non-expanding horizon with null
normal $\ell^a$ such that $[\ca L_{\ell},\widehat{\nabla}_X]=0$ along
$H$.$\blacktriangledown$ 

These horizons are intended to model BHs that are themselves in
equilibrium but possibly in a dynamical space-time. For a detailed
description of their mathematical properties we refer the reader to
Ashtekar-Krishnan's review \cite{Ashtekar:2004cn}. 

The new horizons just introduced all have their own dynamics governed
by Einstein eq.s. There exist for them existence and uniqueness
theorems \cite{Andersson:2005gq}, formulation of first and second laws
\cite{Ashtekar:2004cn,hayw94,Hayward:2004fz} and even a ``membrane
paradigm'' analogy. In
particular, they carry a momentum density which obey a Navier-Stokes
like equation generalizing  the classical Damour's equations of EHs,
except that the bulk viscosity $\zeta_{FOTH}=1/16\pi > 0$
\cite{Gourgoulhon:2005ch,Gourgoulhon:2006uc}. We think Iver would be
amused by that.  

The new horizons are also all space-like or null, hence it remains to
see what is the role they play in the problem of black hole quantum
evaporation. In this connection the following notion can be useful.\\
\noindent \textbf{Time-like Dynamical Horizon:} a smooth three-dimensional, 
\emph{time-like} sub-manifold $H$ of space-time is a time-like dynamical
horizon (TDH) if it can be foliated by closed space-like two-manifolds
$\bm{S}_t$, with future-directed null normals $\ell$ and $n$ such that
(i) on each leaf the expansion $\theta_{(\ell)}$ of one null normal
$\ell^a$ vanishes, (ii) the expansion $\theta_{(n)}$ of the other
null normal $n$ is strictly negative.$\blacktriangledown$

\begin{center}\emph{Surface Gravities}\end{center}

The surface gravity associated to an event horizon is a well known
concept in black hole physics whose importance can be hardly
overestimated. Surprisingly, a number of inequivalent definitions
beyond the historical one 
appeared recently (over the last 15 years or so) in the field with
various underlying motivations. We have collected the following (we
rely on the nice review of Nielsen and Yoon \cite{Nielsen:2007ac}):
\begin{itemize}
\item[1.] The historical Killing surface gravity (Bardeen et
  al.\ \cite{Bardeen:1973gs}, textbooks)  
\item[2.] Hayward's first definition \cite{hayw94}
\item[3.] Mukohyama-Hayward's definition \cite{Mukohyama:1999sp}
\item[4.] Booth-Fairhurst surface gravity for the evolving horizons
  \cite{bf04} 
\item[5.] The effective surface gravity appearing in
  Ashtekar-Krishnan \cite{Ashtekar:2004cn} 
\item[6.] The Fodor et al. definition for dynamical spherically
  symmetric space-times \cite{Fodor:1996rf}
\item[7.] The Visser \cite{Visser:2001kq} and Nielsen-Visser \cite{NV}
  surface  gravity 
\item[8.] Hayward's definition \cite{Hayward:1997jp} using Kodama's
  theory \cite{Kodama:1979vn}.
\end{itemize}
We will not spend much time on the various definitions and their
motivations except for the last item, which is what the tunneling approach
leads to, among other things.

{\bf 1.} The Killing surface gravity is related to the fact that the
integral 
curves of a Killing vector are not affinely parametrized geodesics on
the Killing horizon $H$. Hence
\[
K^a\nabla_bK_a\cong\kappa K_a 
\]
defines the Killing surface gravity $\kappa$ on $H$, where $\cong$
means evaluation on the horizon. The Killing field is supposed to be
normalized at infinity by $K^2=-1$. The definition can be extended to
EHs that are not Killing horizons, by replacing $K$ with the null
generator $\ell$ of the horizon. However there is no preferred
normalization in this case, and this is one reason of the debating
question regarding the value of the surface gravity in dynamical
situations.

{\bf 2.} Hayward's first definition was motivated by the desire to get
a proof 
of the first law for THs. It is defined without appeal to inaffinity
of null geodesics as
\beq\label{hayk1}
\kappa\cong\frac{1}{2}\sqrt{-n^a\nabla_a\theta_{(\ell)}}
\eeq
and is independent on the parametrization of $\ell^a$ integral curves,
since the evaluation is on a marginal outer surface where
$n\cdot\ell=-1$ and $\theta_{(\ell)}=0$.
  
{\bf 3.} We leave apart the Mukohyama-Hayward and  the
Booth-Fairhurst definitions (see {\bf 4.}) 
as they are somewhat more technical and complicated than it is
necessary, so we refer the reader  to the original papers.  

{\bf 5.} Given a weakly isolated horizon $H$, Ashtekar and Krishnan
showed that for any vector 
field $t^a$ along $H$ with respect to which energy fluxes across $H$
are defined, there is an area balance law that takes the form
\[
\de E^t=\frac{\bar{\kappa}}{8\pi G}\de A_S+\mathrm{work\;terms}
\]
with an effective surface gravity given by
\[
\bar{\kappa}=\frac{1}{2R}\frac{dr}{dR}
\]
$R$ is the areal radius of the marginally trapped surfaces,
i.e. $A_S=4\pi R^2$, the function $r$ is related to a choice of
a lapse function and finally $E^t$ is the energy associated with the
evolution vector field $t^a$. For a spherically symmetric DH a natural
choice would be $r=R$ so $\bar{\kappa}=1/2R$, just the result for a
Schwarschild BH. To illustrate the naturalness of this definition,
consider a slowly changing spherically symmetric BH with mass $M(v)$,
where $v$ is a time coordinate. Defining the horizon radius at each
time by $R=2M(v)$ and  $A_S=4\pi R^2$, we can differentiate $M$ to
obtain 
\[
\dot{M}=\frac{\dot{R}}{2}=\frac{1}{2R}\frac{\dot{A_S}}{8\pi}
\;\;\Longrightarrow\;\;\de M=\frac{\bar{\kappa}}{8\pi}\de A_S
\]
which is the usual area balance law with surface gravity
$\bar{\kappa}=1/2R=1/4M$. Consider, however, the more general
possibility where the horizon is at $R=2M(v,R)$, as it happens for
example in the  Vaidya-Bardeen metric. The same computation leads to 
\beq\label{comp}
\dot{M}=\frac{1}{2R}\at1-2M^{'}\ct\frac{\dot{A_S}}{8\pi}\;\;
\Longrightarrow\;\;\kappa\cong \frac{1}{4M}\at1-2M^{'}\ct
\eeq
a prime denoting the radial derivative. Thus naturalness is not a
decisive criterion in this case.

{\bf 6.} The Fodor et al. definition looks like the Killing form of surface
   gravity in that $\kappa\ell^b=\ell^a\nabla_a\ell^b$, where now
   $\ell^a$ is an outgoing null vector orthogonal to a trapped or
   marginally trapped surface. This is because, as a rule, such null
   vectors are not affinely parametrized, although they can always be so
   parametrized that $\kappa=0$. So one needs to fix the
   parametrization: Fodor et al. choose
\[
\kappa=-n^a\ell^b\nabla_b\ell_a
\]
with $n^a$ affinely parametrized and normalized to $n\cdot t=-1$ at
space-like infinity, $t^a$ being the asymptotic Killing field. Note
that this definition is non local but looks like as a natural
generalization of the Killing surface gravity. 

{\bf 7.} We postpone the discussion of the Visser and Visser-Nielsen
surface gravity to the next section.

{\bf 8.} Finally we have a local geometrical definition of this
quantity for the trapping horizon of a spherically symmetric black
hole as \cite{Hayward:1997jp} follows. One
can introduce local null coordinates $x^{\pm}$ in a tubular
neighborhood of a FOTH such that $n=-g^{+-}\partial_-$ and
$\ell=\partial_+$: then (shortening $\theta_{(\ell)}=\theta_+$,
$\theta_{(n)}=\theta_-$)  
\beq\label{kappa}
\kappa=\frac{1}{2}\at g^{+ -}\partial_- \theta_+ 
\ct_{|\theta_+=0}
\eeq 
Later we will show that this $\kappa$ fixes the expansion of the
metric near the trapping outer horizon along a future null
direction. The definition may look somewhat artificial, but in fact it
is very natural and connected directly with what 
is known for the stationary black holes. To see this one notes,
following Kodama \cite{Kodama:1979vn}, that any spherically
symmetric metric admits a
unique (up to normalization) vector field $K^a$ such that $K^aG_{ab}$
is divergence free, where $G_{ab}$ is the Einstein tensor; for
instance, using the double-null form, one finds
$K=-g^{+-}(\partial_+r\partial_--\partial_-r\partial_+)$. The defining
property of $K$ shows that it is a natural
generalization of the time translation Killing field of a static
black hole. Moreover, by Einstein equations $K_aT^{ab}$ will be
conserved so for such metrics there exists a natural localizable
energy flux and its conservation law. Now 
consider the expression $K^a\nabla_{[b}K_{a]}$: it is not hard to see  
that on $H$ it is proportional to $K_b$. The proportionality
factor, a function in fact, is the surface gravity: 
$K^a\nabla_{[b}K_{a]}=-\kappa K_b$. 
For a Killing vector field $\nabla_{b}K_{a}$ is anti-symmetric so the
definition reduces to the usual one.

\section{Two examples: Vaidya and McVittie's metrics}

We consider first spherically symmetric spacetimes which outside the
horizon (if there is one) are described by a metric of the form
\beq
ds^2=-e^{2\Psi(r,v)}A(r,v)dv^2+2 e^{\Psi(r,v)}dvdr+
r^2dS^2\,.
\label{efbard}
\eeq
where the coordinate $r$ is the areal radius commonly used in relation
to spherical symmetry and $v$ is intended to be an advanced null
coordinate. In an asymptotically flat context one can always write (we
use geometrized units in which the Newton constant $G=1$) 
\beq\label{Arv}
A(r,v)=1-2m(r,v)/r
\eeq 
which defines the active mass. This metric was first proposed by Vaidya
\cite{vaidya}, and studied in an interesting paper
during the classical era of black hole physics by Lindquist et al
\cite{lind}. It has been generalized to Einstein-Maxwell systems and
de Sitter space by Bonnor-Vaidya and Mallet, respectively \cite{bvm}. 
It was then extensively used by Bardeen \cite{bardeen}  
and York \cite{york} in their semi-classical analysis of the back 
reaction problem. We will call it the Vaidya-Bardeen metric. A
cosmological constant can be introduced by setting
\beq\label{ArvL}
A(r,v)=1-2m(r,v)/r-r^2/L^2
\eeq
where $L^{-2}\propto\Lambda$. If one wishes the metric can also be
written in double-null form. In the $(v,r)$-plane one can introduce
null coordinates $x^{\pm}$  such that the dynamical Vaidya-Bardeen
space-time may be written as  
\beq
ds^2=-2f(x^+,x^-)dx^+dx^-+r^2(x^+,x^-)dS^2_{D-2}\,,
\label{nf}
\eeq
for some differentiable function $f$. The remaining angular 
coordinates contained in $dS^2$ do not play any essential 
role. In the following we shall use both forms of the metric,
depending on computational convenience. The field equations of
the Vaidya-Bardeen metric are of interest. They
read   
\beq\label{vbe}
\frac{\partial m}{\partial v}=4\pi r^2T^r_{\;v}, \quad \frac{\partial
  m}{\partial r}=-4\pi r^2T^v_{\;v}, \quad 
 \frac{\partial\Psi}{\partial r}=4\pi re^{\Psi}T^v_{\;r}
\eeq
The stress tensor can be written as
\beq\label{st}
T_{ab}=\frac{\dot{m}}{4\pi r^2}\nabla_av\nabla_bv-\frac{m^{'}}{2\pi
  r^2}\nabla_{(a}r\nabla_{b)}v 
\eeq
If $m$ only depends on $v$ it describes a null fluid and obeys the
dominant energy condition if $\dot{m}>0$.

The second example we are interested in is the McVittie solution
\cite{mcv} for a point mass in a Friedmann-Robertson-Walker flat
cosmology. In D-dimensional  spacetime in isotropic spatial
coordinates it is given by \cite{gao} 
\beq
ds^2=-A(\rho,t)dt^2+B(\rho,t)\at d\rho^2+\rho^2dS_{D-2}^2 \ct\,
\eeq
with
\beq
A(\rho,t)=\aq \frac{1-\at \frac{m}{a(t)\rho}\ct^{D-3}}{1+
\at \frac{m}{a(t)\rho}\ct^{D-3}} \cq^2\,,\qquad
 B(\rho,t)=a(t)^2 \aq 1-\at \frac{m}{a(t)\rho}\ct^{D-3}\cq^{2/(D-3)}\,.
\eeq
When the mass parameter $m=0$, it reduces to a spatially flat FRW
solution with scale factor $a(t)$; when $a(t)=1$ it reduces to the
Schwarzschild metric with mass $m$.  
In four dimensions this solution has had a strong impact on the
general problem of matching the Schwarzschild solution with cosmology,
a problem faced also by Einstein and Dirac. 
Besides McVittie, it has been extensively studied by Nolan 
in a series of papers \cite{nolan}. To put the metric in the general
form of Kodama theory, we use what may be called the Nolan gauge, in
which  the metric reads 
\beq\label{nolan}
ds^2=-\at A_s-H^2(t)r^2 \ct dt^2+A_s^{-1}dr^2-2A_s^{-1/2}
H(t)r\,drdt +r^2dS^2_{D-2}
\eeq
where $H(t)=\dot{a}/a$ is the Hubble parameter and, for example, in the
charged 4-dimensional case, 
\beq
A_s=1-2m/r+q^2/r^2
\eeq 
or in $D$ dimension  $A_s=1-2m/r^{D-3}+q^2/r^{2D-6}$. 
 In passing to the Nolan gauge a choice of sign in the cross term
$drdt$ has been done, corresponding to an expanding universe; the
transformation $H(t)\to-H(t)$ changes this into a contracting one. 
In the following we shall consider $D=4$ and $q=0$; then the
Einstein-Friedmann equations  read
\beq\label{eeqs}
3H^2=8\pi\rho\,, \qquad 2A_s^{-1/2}\dot H(t)+3H^2=-8\pi p\,.
\eeq
It follows that $A_s=0$, or $r=2m$, is a curvature singularity. In
fact, it plays the role that 
$r=0$ has in FRW models,  namely it is a big bang
singularity.  
When $H=0$ one has the Schwarzschild solution. Note how the term
$H^2r^2$ in the metric strongly resembles a varying cosmological
constant; in fact for $H$ a 
constant, it reduces to the Schwarzschild-de Sitter solution in Painlev\'e
coordinates. As we will see, the McVittie solution possesses in general 
both apparent and trapping horizons, and the spacetime is
dynamical. However, it is really not a dynamical black hole in the
sense we used it above, since the mass parameter is strictly constant:
for this reason we called it a fake dynamical BH. This observation
prompts one immediately for an obvious extension of the solution: to
replace the mass parameter by a function of time and radius, but this
will not be pursued here.  

The study of black holes requires  also a notion of energy; the natural
choice would be to use the charge associated to Kodama conservation
law, but this turns out to be the Misner-Sharp energy, which for a
sphere with areal radius $r$ is the same as the Hawking
mass \cite{hawk}, given by $E=r(1-2^{-1}r^2g^{+-}\theta_+\theta_-)/2$.   
Using the metric \eqref{efbard} an equivalent expression is
\beq\label{hmass}
g^{\mu\nu}\partial_{\mu}r\partial_{\nu}r=1-2E/r
\eeq
In this form it is clearly a generalization of the Schwarzschild
mass. As we said, $E$ is just the charge associated to Kodama conservation
law; as showed by Hayward \cite{Hayward:1994bu}, in vacuum $E$ is
also the Schwarzschild energy, at null infinity it is the Bondi-Sachs
energy and at spatial infinity it reduces to the ADM mass.

Let us apply this general theory to the two classes of dynamical BH we
have considered. Using Eq.~\eqref{efbard}, we have 
$\theta_{(\ell)}=A(r,v)/2r\,.$ The condition $\theta_{(\ell)}=0$
leads to $A(r_h,v)=0$,  
which defines a curve $r_h=r_h(v)$ giving the location of the 
horizon; it is easy to show that $\theta_-<0$, hence the horizon is of
the future type. Writing the solution in the Vaidya-Bardeen 
form, that is with $A(r,v)=1-2m(r,v)/r$, the Misner-Sharp energy of the
black hole is $E=m(r_h(v),v)$, and the horizon will be outer trapping if
$m^{'}(r_h,v)<1/2$, a prime denoting the radial derivative.
The geometrical surface gravity associated with the Vaidya-Bardeen
dynamical horizon is
\beq\label{vbk}
\kappa(v)\cong\frac{A^{'}(r,v)}{2}=\frac{m(r_h,v)}{r_h^2}-
\frac{m^{'}(r_h,v)}{r_h}=\frac{1}{4m}\at1-2m^{'}\ct
\eeq
the same as Eq.~\eqref{comp}, where $m\equiv m(r_h,v)$. We see the
meaning of the outer trapping condition: it ensures the positivity of
the surface gravity.  

As a comparison, Hayward's first definition would give
$\kappa=\sqrt{1-2m^{'}}/4m$, while Fodor et al. expression is 
\beq\label{fodor}
\kappa=\frac{2^{\Psi}}{4m}(1-2m^{'})+\dot{\Psi}
\eeq
The effective surface gravity of Ashtekar-Krishnan simply is
$\kappa=1/4m$, everything being evaluated on the horizon. 
Note that some of them are not correct for the Reissner-Nordstr\"{o}m
black hole. We also note that $\kappa$ (\ref{vbk}) is inequivalent to
the Nielsen-Visser surface gravity, which in these
coordinates takes the form
\beq
\tilde\kappa=\frac{1}{4m}\,(1-2m^{'}-e^{-\Psi}\dot{m})
\eeq
though they  coincide in the static case. Also, both are inequivalent to the
Visser surface  gravity $e^\Psi\tilde\kappa$,
which was derived as a  temperature by essentially the same tunneling
method as discussed below, but in Painlev\'e-Gullstrand
coordinates. Part of the difference can be traced to a different choice
of time. 

In the case of McVittie BHs, we
obtain 
\[
\theta_{\pm}=\pm(\sqrt A_s\mp H r)/2rf_{\pm}
\] 
where the functions $f_{\pm}$ are integrating factors determining null
coordinates $x^{\pm}$ 
such that $dx^{\pm}=f_{\pm} \aq \at \sqrt A_s\pm H r \ct dt\pm
A_s^{-1/2}dr\cq$.  One may compute from this the dual derivative
fields $\partial_{\pm}$.  
The future dynamical horizon  defined by $\theta_+=0$, has a
radius which is a root of the equation $\sqrt A_s=H r_h \,,$
which in turn implies $A_s=H^2r^2_h$. Hence the horizon radius is a
function of time. The Misner-Sharp mass and the related surface
gravity are 
\beq\label{msmcv}
E=m+\frac{1}{2}\,H(t)^2r_h^3
\eeq
\beq
\kappa(t)=\frac{m}{r_h^2}-H^2 r_h -\frac{ \dot H}{2 H}
=\frac{E}{r_h^2}-\frac{3}{2}\,H^2 r_h -\frac{ \dot H}{2 H}  
\label{mvsg}
\eeq
Note that $E=r_h/2$. In the static cases everything agrees with the
standard  results. The 
 surface gravity has an interesting expression in terms of the sources
of Einstein equations and the Misner-Sharp mass. Let $\tilde T$ be the
reduced trace of the stress tensor in the space normal to the sphere
of symmetry, evaluated on the horizon $H$. For the Vaidya-Bardeen
metric it is, by Einstein's equations \eqref{vbe},
\[
\tilde T=T^v_{\;v}+T^r_{\;r}=-\frac{1}{2\pi r_h}\,\frac{\partial
  m}{\partial r}_{|r=r_h}
\]
For the McVittie's solution, this time by Friedmann's equations
\eqref{eeqs} one has  
\[
\tilde T=-\rho+p=-\frac{1}{4\pi}\at 3H^2+\frac{\dot{H}}{Hr_h}\ct
\]  
We have then the mass formula
\beq\label{mfor}
\frac{\kappa A_H}{4\pi} = E+2\pi r^3_h\tilde T
\eeq 
where $A_H=4\pi r_h^2$. It is worth mentioning the pure FRW case,
i.e. $A_s=1$, for which $\kappa(t)= -\at H(t)+\dot H/2 H\ct\,.$ One
can easily see that \eqref{mfor} is fully equivalent to Friedmann's
equation.   
We feel that these expressions for the surface gravity are non trivial
and display deep connections with the emission process. Indeed it is
the non vanishing of $\kappa$ that is connected with the imaginary
part of the action of a massless particle, as we are going to show in
the next section. 

\section{Tunneling and instability} 

The essential property of the tunneling method is that the action
$I$ of an outgoing massless particle emitted from the horizon has an
imaginary part which for stationary black holes is $\Im
I=\pi\kappa^{-1}E$, where $E$ is the Killing energy and $\kappa$ the 
horizon surface gravity. The imaginary part is obtained by means of
Feynman $i\epsilon$-prescription, as explained in
\cite{parikh,angh}. As a result the particle production rate reads  
$\Gamma= \exp(-2 \Im I)=\exp(-2\pi\kappa^{-1}E)\,.$
One then recognizes the Boltzmann factor, from which one deduces the
well-known temperature $T_H=\kappa/2\pi$. Moreover, an explicit
expression for $\kappa$ is actually obtained in terms of radial
derivatives of the metric on the horizon.

Let us consider now the case of a dynamical black hole in the double-null  
form \cite{Di Criscienzo:2007fm}. We have for a massless particle
along a radial geodesic the 
Hamilton-Jacobi equation $\partial_+I \partial_-I=0\,.$ 
Since the particle is outgoing $\partial_- I$ is not vanishing, and we
arrive at the simpler condition  $\partial_+ I=0$. 
First, let us apply this condition to the Vaidya-Bardeen BH.  One
has then
\beq\label{hjv}
2e^{-\Psi(r,v)}\partial_v I+A(r,v)\,\partial_r I=0\,.
\eeq
Since the particle will move along a future null geodesic, to pick
the imaginary part  we expand the metric along a future null direction
starting from an arbitrary event $(r_h(v_0),v_0)$ on the horizon,
i.e. $A(r_h(v_0),v_0)=0$.  Thus, shortening $r_h(v_0)=r_0$,
we have 
$A(r,v)=\partial_r A(r_0,v_0)\Delta r+\partial_vA(r_0,v_0)\Delta
v+\dots=2\kappa(v_0)(r-r_0)+\dots$,  
since along a null direction at the horizon $\Delta v=0$, according to 
 the metric \eqref{efbard}; here $\kappa(v_0)$ is the surface gravity, 
Eq.~\eqref{vbk}. 
From \eqref{hjv} and the expansion, $\partial_rI$ has a simple pole at
the event $(r_0,v_0)$; as a consequence
\beq
\Im I=\Im \int  \partial_r I dr=-\Im  \int dr
\frac{2e^{-\Psi(r,v)}\partial_v I}{A^{'}(r_0,v_0)(r-r_0-i0)} 
=\frac{\pi \omega(v_0) }{\kappa(v_0)}\,. 
\eeq
where  $\omega(v_0)= e^{-\Psi(r_0,v_0)}\partial_vI $, is to be
identified with  the energy of the particle at the time $v_0$. Note that
the Vaidya-Bardeen metric has a sort of gauge invariance due to conformal
reparametrizations of the null coordinate $v$: the map $v\to\tilde
v(v)$, $\Psi(v,r)\to\tilde\Psi(\tilde v,r)+\ln(\partial\tilde v/\partial
v)$ leaves the metric invariant, and the energy is gauge invariant
too. Thus we see that the Hayward-Kodama surface gravity appears to be 
relevant to the process of particles emission. The emission
probability, $\Gamma= \exp(-2\pi\omega(v)/\kappa(v))$, has the form of
a Boltzmann factor, suggesting a locally thermal spectrum.

For the McVittie's BH, the situation is similar. In fact, the
condition $\partial_+I=0$ becomes 
\[
\partial_rI=-F(r,t)^{-1}\partial_t I
\]
where 
\[
F(r,t)=\sqrt{A_s(r)}\at\sqrt{A_s(r)}-rH(t)\ct
\]
As before, we pick
the imaginary part by expanding this function at the  horizon  
along a future null direction, using the fact that for two
neighbouring events on a null direction in the metric \eqref{nolan},
one has $t-t_0=(2H_0^2r_0^2)^{-1}(r-r_0)$, where $H_0=H(t_0)$. We find
the result 
\beq\label{exp1}
F(r,t)=\at\frac{1}{2}\,A^{'}_s(r_0)-r_0 H_0^2-\frac{\dot H_0}{2H_0}\ct
(r-r_0)+\dots = \kappa(t_0)(r-r_0)\dots 
\eeq
where this time $r_0=r_h(t_0)$.
From this equation we see that $\partial_rI$ has a simple pole at the
horizon; hence, making use again of Feynman $i\epsilon$-prescription,
one finds $\Im I=\pi\kappa(t_0)^{-1}\omega(t_0)$, where
$\omega(t)=\partial_tI$ is again the energy at time $t$,   
in complete agreement with the geometric evaluation of the previous
section. Obviously, if $\kappa$ vanishes on the horizon there is no
simple pole and the black hole should be stable\footnote{However,
  charged  extremal black holes can radiate \cite{Vanzo:1995bh}.}. The
kind of instability producing the Hawking flux for 
stationary black holes evidently persists in the dynamical arena, and
so long as the evolution is sufficiently slow the black hole seems to
evaporate thermally. Note that the imaginary part, that is the
instability, is attached to the horizon all the time, confirming the
Fredenhagen-Haag suggestion quoted in the introduction. It is worth
mentioning the role of $\kappa$ in the analogue of the first law for
dynamical black holes (contributions to this problem for Vaidya black
holes were given in \cite{cinesi}). Using the formulas of the
projected stress tensor $\tilde T$ given above, and the expression of
the Misner-Sharp energy, one obtains the differential law 
\beq\label{fl}
dE=\frac{\kappa\, dA_H}{8\pi}-\frac{\tilde T}{2}\,dV_H
\eeq
provided all quantities were computed on the trapping horizon.
Here $A_H=4\pi r_H^2$ is the horizon area and $V_H=4\pi r_H^3/3$ is a
formal horizon volume. If one interprets the ``d'' operator as
a derivative along the future null direction one gets Hayward's form
of the first law. But one can also interpret the differential
operation more abstractly, as referring to an ensemble. Indeed, to
obtain Eq.~\eqref{fl} it is not necessary to specify the meaning of the
``d''. It is to be noted that the same law can be proved with other,
inequivalent definitions of the surface gravity, even maintaining the
same meaning for the energy. Thus other considerations are needed to
identify one: the tunneling method has made one choice. 

As thoroughly discussed in Hayward et al.\ \cite{srmls},
Eq.~\eqref{efbard} is actually the most general form of a spherically 
symmetric metric, so the above calculations works throughout. Of course
$\kappa >0$ if the trapping horizon is of the outer type.
Thus the method has derived a positive temperature if and only if there 
is a future outer trapping horizon.  

\begin{center}\emph{Extremal limit}\end{center}

We discuss only an example, the charged stringy black hole, 
which represents a non-vacuum solution of Einstein-Maxwell dilaton gravity in 
the string frame \cite{gibbons,garf}: 
\begin{equation} ds^2=r^2d \Omega^2 
+\frac{dr^2}{\left(1-a/r\right) \left(1-b/r\right)} 
-\left( \frac{1-a/r}{1-b/r}\right)dt^2 
\end{equation} 
where $a>b>0$. The horizon radius is $r = a$. 

For this example, the extremal limit as defined by global structure is $b 
\rightarrow a $. The Killing surface gravity $\kappa_{\infty}\cong1/2a$ 
does not vanish in this limit. Garfinkle et al.\ \cite{garf} noted this as 
puzzling, since extremal black holes are expected to be zero-temperature 
objects. 

Remarkably, the geometrical surface gravity \eqref{vbk} 
\begin{equation} 
\kappa\cong\frac{a-b}{2 a^2} 
\end{equation} 
vanishes in the extremal limit. Thus the gravitational dressing effect lowers 
the temperature to its theoretically expected value. 

We conjecture that this is  true in general. Indeed, past experience with 
extremal black holes showed that the horizon of these objects is not only a 
zero but also a minimum of the expansion $\theta_+=\partial_+A/A$ of the 
radially outgoing null geodesics, $\theta_+$ becoming positive again on 
crossing the horizon. Thus $\partial_-\theta_+\cong0$ should be the
appropriate  definition of an extremal black hole. Since 
$\kappa=-e^{-2\varphi}\partial_-\partial_+r$, this is equivalent to 
$\kappa = 0$.

\end{document}